\documentclass[twocolumn,prb,showpacs,floatfix,superscriptaddress]{revtex4-1}

\usepackage{graphicx}
\usepackage{bm}
\usepackage{amsmath}
\usepackage{amsfonts}
\usepackage{amssymb}
\usepackage{xcolor}
\usepackage{float}
\usepackage{xspace}
\usepackage{scalerel}

\newcommand{\cm}{~cm$^{-1}$}
\newcommand{\aIO}{$\alpha$-A$_{2}$IrO$_{3}$ (A=Na, Li)\xspace}
\newcommand{\aNIO}{Na$_2$IrO$_3$\xspace}
\newcommand{\aLIO}{$\alpha$-Li$_2$IrO$_3$\xspace}
\newcommand{\bLIO}{$\beta$-Li$_2$IrO$_3$\xspace}
\newcommand{\gLIO}{$\gamma$-Li$_2$IrO$_3$\xspace}
\newcommand{\aRC}{$\alpha$-RuCl$_3$\xspace}
\newcommand{\LIO}{$\alpha, \beta, \gamma$-Li$_2$IrO$_3$\xspace}
\newcommand{\sio}{Sr$_2$IrO$_4$\xspace}

\begin{document}
\title{\textbf{Lattice dynamics and structural transition of the hyperhoneycomb iridate \bLIO investigated by high-pressure Raman scattering}}

\author{Sungkyun Choi}
\altaffiliation[Present address: ] {Department of Physics and Astronomy, Rutgers University, Piscataway, New Jersey 08854, USA}
\email{sc1853@physics.rutgers.edu}
\affiliation{Max Planck Institute for Solid State Research, Heisenbergstrasse 1, 70569 Stuttgart, Germany}

\author{Heung-Sik Kim}
\affiliation{Department of Physics and Astronomy, Rutgers University, Piscataway, New Jersey 08854-8019, USA}
\affiliation{Department of Physics, Kangwon National University, 1 Gangwondaehak-gil, Chuncheon-si, Gangwon-do 24341, Republic of Korea}

\author{Hun-Ho Kim}
\affiliation{Max Planck Institute for Solid State Research, Heisenbergstrasse 1, 70569 Stuttgart, Germany}

\author{Aleksandra Krajewska}
\affiliation{Max Planck Institute for Solid State Research, Heisenbergstrasse 1, 70569 Stuttgart, Germany}

\author{Gideok Kim}
\affiliation{Max Planck Institute for Solid State Research, Heisenbergstrasse 1, 70569 Stuttgart, Germany}

\author{Matteo Minola}
\affiliation{Max Planck Institute for Solid State Research, Heisenbergstrasse 1, 70569 Stuttgart, Germany}

\author{Tomohiro Takayama}
\affiliation{Max Planck Institute for Solid State Research, Heisenbergstrasse 1, 70569 Stuttgart, Germany}

\author{Hidenori Takagi}
\affiliation{Max Planck Institute for Solid State Research, Heisenbergstrasse 1, 70569 Stuttgart, Germany}

\author{Kristjan Haule}
\affiliation{Department of Physics and Astronomy, Rutgers University, Piscataway, New Jersey 08854-8019, USA}

\author{David Vanderbilt}
\affiliation{Department of Physics and Astronomy, Rutgers University, Piscataway, New Jersey 08854-8019, USA}

\author{Bernhard Keimer}
\affiliation{Max Planck Institute for Solid State Research, Heisenbergstrasse 1, 70569 Stuttgart, Germany}

\date{\today}

\begin{abstract}
We report a polarized Raman scattering study of the lattice dynamics of \bLIO~under hydrostatic pressures up to 7.62~GPa. At ambient pressure, \bLIO~exhibits the hyperhoneycomb crystal structure and a magnetically ordered state of spin-orbit entangled $J_{\rm eff}$ = 1/2 moments that is strongly influenced by bond-directional (Kitaev) exchange interactions. At a critical pressure of $\sim 4.1$~GPa, the phonon spectrum changes abruptly consistent with the reported structural transition into a monoclinic, dimerized phase. A comparison to the phonon spectra obtained from density functional calculations shows reasonable overall agreement. The calculations also indicate that the high-pressure phase is a nonmagnetic insulator driven by the formation of Ir--Ir dimer bonds. Our results thus indicate a strong sensitivity of the electronic properties of \bLIO~to the pressure-induced structural transition.
\end{abstract}
\maketitle
\noindent

\section{Introduction}
The effect of spin-orbit coupling (SOC) on the electronic structure of heavy transition metal compounds with 4$d$ and 5$d$ valence electrons has attracted great attention, especially in search and characterization of unprecedented electronic phases and their dynamics.~\cite{witczak} A prominent example is the Kitaev quantum spin liquid, which exhibits unconventional quantum entanglement and fractionalized excitations, in contrast to conventional magnetic ordering phenomena.~\cite{kitaev}

The search for a physical realization of the Kitaev spin liquid has motivated an intense research effort on honeycomb-based lattices with edge-sharing IrO$_{6}$ (Ir$^{4+}$) or RuCl$_{6}$ (Ru$^{3+}$) octahedra. The strong SOC of the Ir or Ru ions gives rise to the formation of local $J_{\rm eff}=1/2$ moments and to bond-dependent Kitaev exchange interactions.~\cite{jackeli-bonds,Jackeli2010} Candidates for Kitaev magnetism have also been identified in materials with three-dimensional lattice architectures.~\cite{Mandal2009,Lee2014,Itamar2015}

\begin{figure*}
\begin{center}
\includegraphics[width=\linewidth]{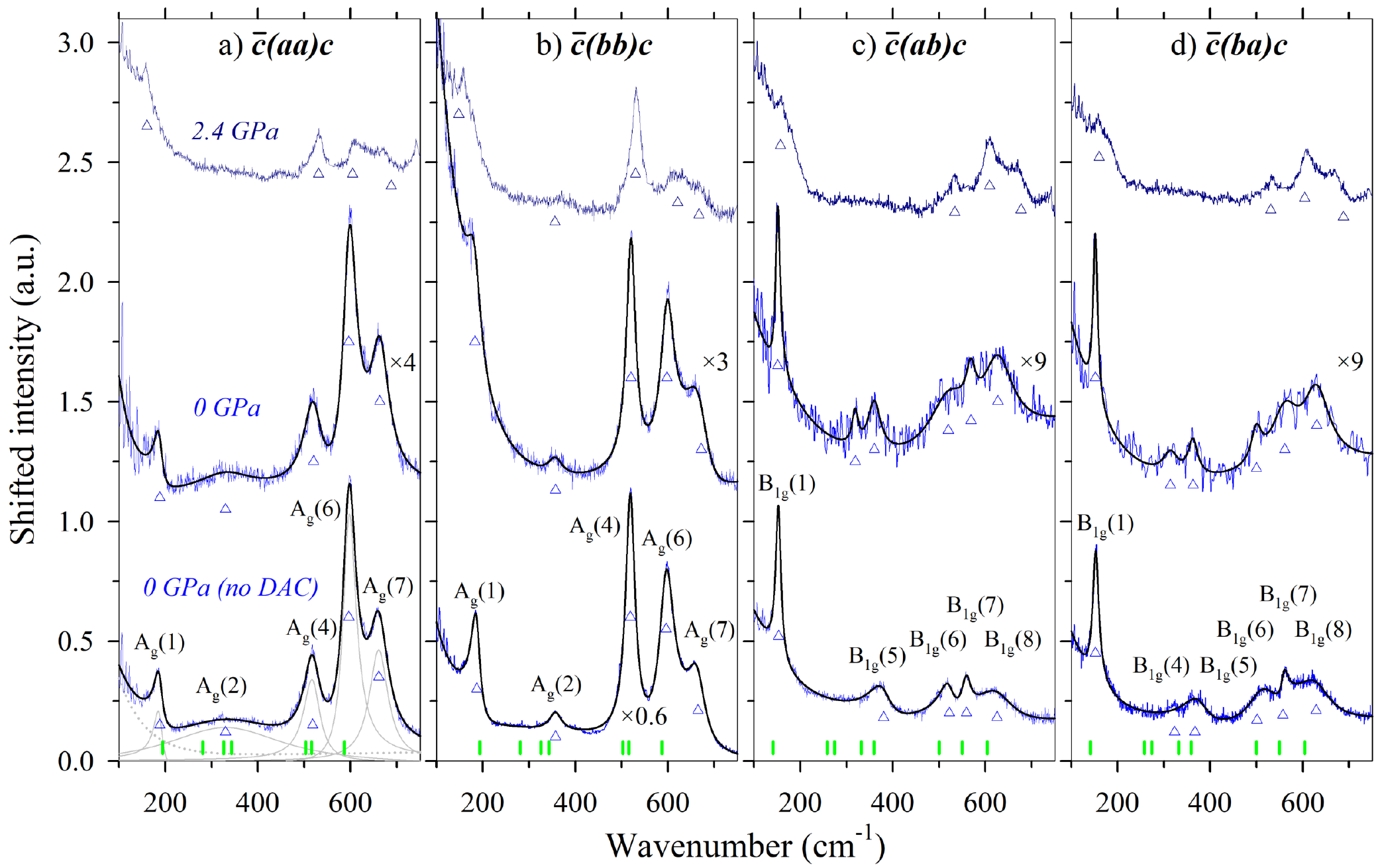}
\end{center}
\caption {Polarized Raman spectra at 0~GPa and 2.4~GPa measured with the green laser. Four polarization channels are shown: a) $\bar{\bm c}(\bm a \bm a){\bm c}$, b) $\bar{\bm c}(\bm b \bm b){\bm c}$, c) $\bar{\bm c}(\bm a \bm b){\bm c}$ and d) $\bar{\bm c}(\bm b \bm a){\bm c}$ defined at ambient pressure. Two ambient-pressure data sets (with and without diamond anvil cell (DAC)) are shown. Blue lines are data and solid black lines are fitted curves (with Fano profiles for the phonon modes). Upward triangular symbols indicate the peak positions obtained from the fits. Gray solid lines for the 0~GPa data (without DAC) in a) are decomposed contributions of each peak illustrated as a representative fit. The dotted line is the fitted background signal. The amplified 0~GPa data and the raw 2.4~GPa data were shifted vertically for more direct comparison. Calculated frequencies from density functional calculations are given with green vertical bars for comparison with the experimental data.
}
\label{fig:GR:DAC}
\end{figure*}

However, almost all known Kitaev-candidate materials -- \aIO,~\cite{Gegenwart2010} \bLIO,~\cite{Takayama:beta:Li213} \gLIO,~\cite{Modic2014} and~\aRC~\cite{Plumb2014} -- appear to show long-range magnetic order (at least in the absence of external magnetic fields): zigzag antiferromagnetism for Na$_2$IrO$_3$~\cite{Liu2011,Na213_INS_2012, Ye2012, Hwan2015} and $\alpha$-RuCl$_3$,~\cite{Johnson2015,Banerjee2016,Do2017} and incommensurate counterrotating magnetic order for the Li$_2$IrO$_3$ family.~\cite{Steph:Li213,alun_beta_Li213,alun_gamma_Li213,Choi:Li213} The appearance of these long-range magnetic orders is currently understood as a consequence of the presence of non-negligible additional exchange interactions such as second- or third-nearest-neighbor Heisenberg terms~\cite{Itamar2015} or symmetric anisotropy interactions~\cite{Eric2016} in addition to the predominant Kitaev interaction.

To study and manipulate the subdominant exchange couplings and move closer to the realization of a Kitaev quantum spin liquid, a number of recent investigations has focused on the influence of lattice distortions (especially a trigonal distortion in honeycomb-lattice materials~\cite{Stephen2017,Katukuri2014}). To this end, both external pressure and chemical methods such as hydrogen intercalation have been used to modify the lattice structure of various Kitaev materials.~\cite{Hermann2017,Simutis2018,Hermann2018,Clancy2018,Veiga2017,Majumder2018, Biesner2018,Garbarino2018,Raman:Rucl3,Kitagawa2018,Hermann2019,Veiga2019,Breznay2017,Yadav2018}

In particular, a recent high-pressure study of \bLIO~with x-ray diffraction found a signature of a structural transition from the orthorhombic structure at ambient pressure ($Fddd$ space group, No.~70, $mmm$ point group)~\cite{Takayama:beta:Li213,alun_beta_Li213} to a lower-symmetry monoclinic structure ($C2/c$ space group, No.~15, $2/m$ point group)~\cite{Veiga2017} around P $\sim$ 4~GPa at room temperature. This study was later extended to low temperatures.~\cite{Veiga2019} Recently, similar structural transitions have been experimentally reported in \aRC~\cite{Biesner2018} above 1~GPa, in \aLIO~\cite{Clancy2018,Hermann2018,Hermann2019} around 3.8~GPa, and also theoretically predicted in \aNIO around 36~GPa~\cite{Hu2018} (experimentally no structural transition was observed below about 25~GPa~\cite{Hermann2017}).

Investigations of the lattice dynamics under pressure yield information complementary to diffraction techniques, and can potentially provide insight into static and dynamic spin-lattice coupling. Optical spectroscopy has been recently used to study this relationship in several materials including \aNIO,~\cite{Hermann2017} \aLIO,~\cite{Hermann2019} and \aRC.~\cite{Biesner2018}

However, pressure-dependent Raman scattering studies of Kitaev materials have rarely been reported; only \aRC has been examined.~\cite{Raman:Rucl3} Having distinct selection rules for phonons, Raman scattering is a suitable tool for this purpose, with a high sensitivity to small structural modifications.~\cite{Taheri2016:Han2018} Moreover, it can capture Raman-active optical phonons with high energy resolution, which makes detailed analysis of the phonon energies possible even under pressure.

Among Kitaev materials, the \bLIO~\cite{Takayama:beta:Li213} compound can be a good choice due to its three-dimensional Ir network that is less prone to structural defects common in layered compounds such as \aIO~\cite{Na213_INS_2012,Freund2016} and \aRC.~\cite{Johnson2015} In addition, because of its more ideal IrO$_{6}$ octahedral structure, \bLIO~is expected to be closer to the Kitaev spin liquid than its structural analogue \gLIO.~\cite{Modic2014}

Here we have confirmed the existence of the recently reported structural transition by high-pressure Raman measurements on \bLIO~single crystals. We clearly observed the splitting and broadening of Raman-active phonon peaks and the development of multiple new Raman modes at high pressure, which are Raman hallmarks of a first-order structural transition to a lower crystal symmetry. At ambient pressure, polarization analysis allowed us to distinguish different Raman modes based on the Raman selection rules of a given crystal symmetry. The measured frequencies of Raman-active phonons both at ambient and high pressure were compared to those from {\it ab initio} density functional theory (DFT) and dynamical mean-field theory calculations. Our combined analysis suggests that the lower-symmetry monoclinic phase at high pressure originates from the dimerization of Ir ions, transforming the Ir atomic 5$d$ orbitals into bonding and antibonding dimer states. This phase does not accommodate local $J_{\rm eff}$ = 1/2 moments, indicating a delicate balance between magnetism and the intermetallic covalency. These conclusions are also consistent with a very recent neutron and resonant inelastic x-ray scattering study~\cite{Takayama2019} characterizing the pressure-induced structural transition at room temperature.

The paper is organized as follows. Sections~\ref{sec:exp} and~\ref{sec:comp} describe details of the Raman scattering measurements and {\it ab initio} calculations, respectively, followed by results and discussions in Sec.~\ref{sec:hpRaman}. Phonon spectra from high-pressure polarization-resolved Raman measurements on \bLIO~single crystals are presented in Sec.~\ref{sec:hpRaman:green}. We then present and discuss computational results in Sec.~\ref{sec:abinitio}, followed by comparison between experimental and computational data in Sec.~\ref{sec:hpRaman:phonon}. We summarize our conclusions in Sec.~\ref{sec:summary}.

\section{Experimental Details}
\label{sec:exp}
\begin{figure*}[tbh]
\includegraphics[width=\linewidth]{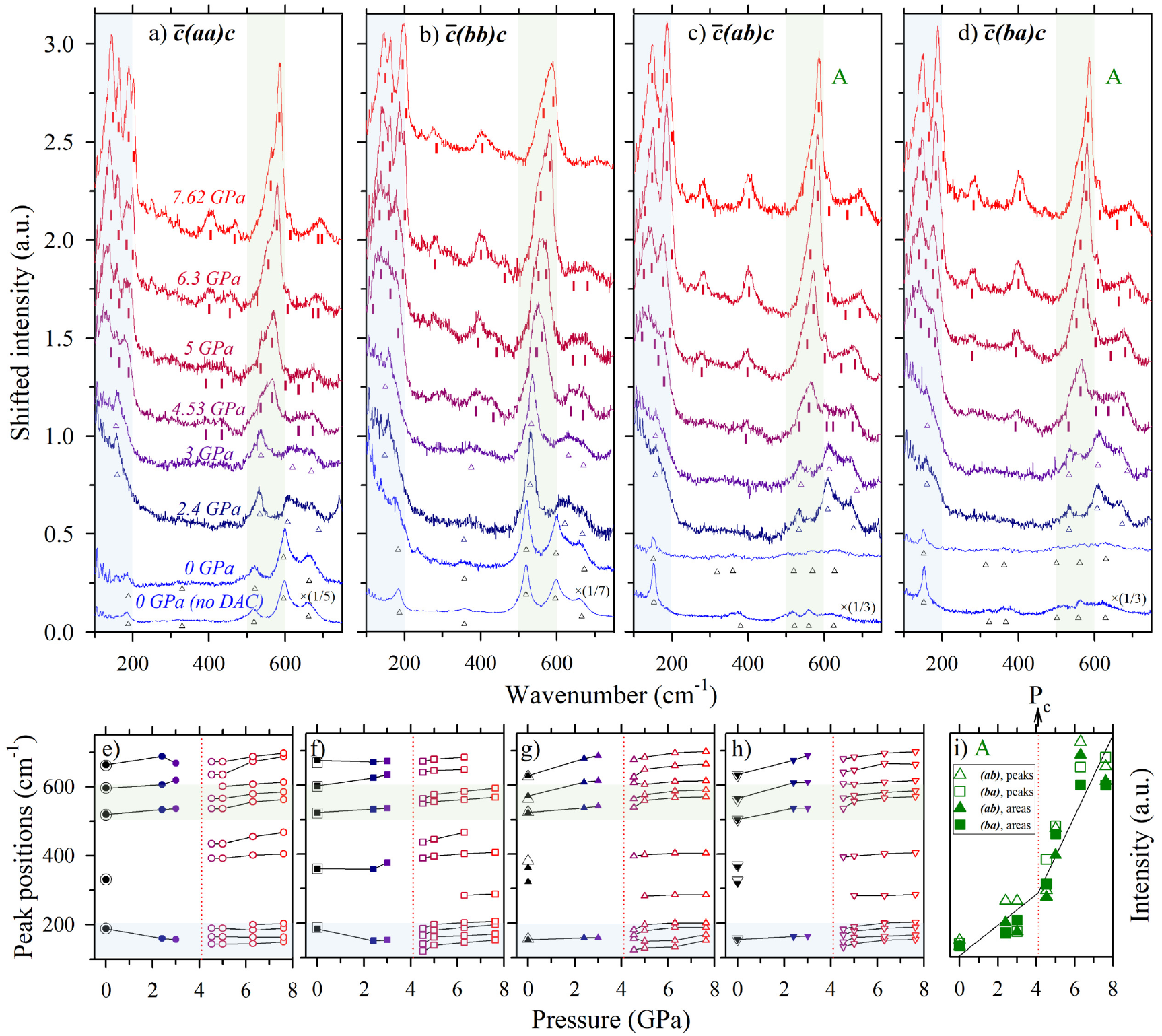}
\caption {Polarized Raman data taken with the green laser as a function of pressure in four polarization geometries: a) $\bar{\bm c}(\bm a \bm a){\bm c}$, b) $\bar{\bm c}(\bm b \bm b){\bm c}$, c) $\bar{\bm c}(\bm a \bm b){\bm c}$ and d) $\bar{\bm c}(\bm b \bm a){\bm c}$ defined at ambient pressure, corresponding to A$_{\rm g}$, A$_{\rm g}$, B$_{\rm 1g}$ and B$_{\rm 1g}$ Raman-active modes, respectively. In the monoclinic structure at high pressure (4.53~GPa and above), only parallel Raman-polarized modes (A$^{*}_{\rm g}$) were observed owing to the ($\bm{ab}$)-plane-oriented mounting of the four polarization setups in this experiment (see the text for the definition of the asterisk symbol). We confirmed that in-plane rotation of the sample did not noticeably affect the Raman spectra at ambient pressure~\cite{SM}. Triangular (vertical bar) symbols are the extracted peak positions below (above) the critical pressure for the structural transition. Two spectral ranges with the most dramatic change of the Raman spectra with pressure are emphasized with transparent blue and green shaded areas at around 150\cm~and 550\cm, respectively. The 0~GPa data without the DAC were scaled and all other data were vertically shifted for better comparison. e-h) Evolution of peak positions with pressure obtained from a-d). Bigger empty symbols from the 0~GPa data without the DAC are compared with solid symbols from the 0~GPa data with the DAC, confirming an excellent match of phonon frequencies. i) Two representations of summed intensities of Raman data highlighted as green boxes in c-d): summed peak intensities between 500 and 600\cm~(empty symbols) and integrated areas between 550 and 600\cm~(filled symbols). The solid black line is a guide to the eyes and the vertically dotted red line marks the estimated critical pressure at about 4.1~GPa, which is also marked in e-h). }
\label{fig:GR}
\end{figure*}

Single crystals of \bLIO~were grown by a flux method~\cite{Takayama:beta:Li213}. We measured Raman spectra on more than 40 crystals, which were consistent with the previous result at ambient pressure.~\cite{Glamazda2016} We then screened crystals in terms of better signal-to-noise ratio, clearer surface morphology, and shinier surface to proceed with high-pressure Raman measurements. Raman data were also acquired from both green and red laser sources, revealing that the dominant Raman phonons in the spectra using incident green laser (514.5 nm) were stronger than those collected with the incident red laser (632.8 nm). The complete polarized Raman measurements were therefore carried out with the green laser [see Fig.~\ref{fig:GR:DAC} and ~\ref{fig:GR}], complemented by measurements with the red laser [see Fig.~\ref{fig:RR:pol}]. The former were fitted by Fano profiles~\cite{Fano1961} to extract the peak positions.

All measurements used a backscattering configuration and hereafter we use Porto's notation~\cite{Damen1966} to specify the experimental geometry.~\cite{SM} With the backscattering geometry, we employed $\bar{\bm c}(\bm a \bm a){\bm c}$, $\bar{\bm c}(\bm b \bm b){\bm c}$, $\bar{\bm c}(\bm a \bm b){\bm c}$, $\bar{\bm c}(\bm b \bm a){\bm c}$ configurations to probe A$_{\rm g}$, A$_{\rm g}$, B$_{\rm 1g}$ and B$_{\rm 1g}$ modes at ambient pressure~\cite{Glamazda2016}, where $\bm{a}$, $\bm{b}$ and $\bm{c}$ are the orthorhombic crystallographic axes. In the monoclinic structure at high pressure, all polarization geometries used in this study can only probe A$^{*}_{g}$ phonons due to different Raman selection rules~\cite{SM} (Table~\ref{table:PP}) from the orthorhombic symmetry at ambient pressure. Note that the asterisk (*) symbol is included to indicate that the experimentally measured A$_{\rm g}$ phonons were not obtained from the exact backscattering condition at high pressures due to an inclined $\bm{c}$-axis in the monoclinic structure ($\beta$=106.777$^{\circ}$)~\cite{Veiga2017} relative to the backscattering direction used in the experimental setup (see Table~\ref{table:PP} and Supplemental Material~\cite{SM} for the full details).

High-pressure Raman measurements with both laser lines were conducted with a mechanically driven gasketed diamond anvil cell (Stuttgart type). A crystal that had been characterized by Raman scattering at ambient pressure was placed inside the hole of the gasket with a 4~:~1 methanol-ethanol liquid as a pressure medium to ensure hydrostatic pressure conditions up to 10.5~GPa.~\cite{Met_Eth} Potential difficulties arising from increased viscosity of the pressure medium were avoided by keeping the pressure below 7.63~GPa. Pressures were measured by the ruby luminescence method with four ruby balls~\cite{Ruby:HP} spread spatially next to a \bLIO~crystal inside the gasket to accurately evaluate hydrostatic pressures, and were repeated before and after collecting the Raman data at each pressure, confirming only a small variation of pressure ($\Delta P \lesssim$ 0.1~GPa). Raman spectra was also reproduced after releasing pressure (not shown) and a similar transition pressure and Raman spectra were found, indicating that this structural transition was reversible within the pressure explored. A typical collection time was about 10 hours with the DAC to maximize the signal to noise ratio.

\section{Computational Details}
\label{sec:comp}
In the {\it ab initio} calculation, we used density functional theory (DFT) augmented by atomic SOC and on-site Coulomb interactions (DFT + SOC +$U$) where $U$ indicates the electronic correlation. The Vienna {\it ab initio} Simulation Package ({\sc vasp})~\cite{Kresse1996,Kresse1999} was employed, supplemented by {\sc wien2k}~\cite{wien2k} and DFT+embedded DMFT Functional code (eDMFT)~\cite{eDMFT,Haule2010,Haule2018} calculations. Note that DFT or DFT+SOC was also employed when needed. The {\sc phonopy} package was employed to calculate the $\Gamma$-point phonon modes based on the relaxed orthorhombic and monoclinic structures.~\cite{Togo2015}

\section{Results and discussion}
\label{sec:hpRaman}

\subsection{Raman experiments}
\label{sec:hpRaman:green}
Figure~\ref{fig:GR:DAC} presents Raman data on \bLIO~\cite{Glamazda2016} with phonon peaks identified. A group-theoretical analysis of the space group $Fddd$ reveals the following irreducible representations: $\Gamma$ = 7 A$_{\rm g}$ $(\bm a \bm a, \bm b \bm b, \bm c \bm c)$ + 8 B$_{\rm 1g}$ $(\bm a \bm b)$ + 11 B$_{\rm 2g}$ $(\bm a \bm c)$ + 10 B$_{3g}$ $(\bm b \bm c)$~\cite{Glamazda2016}. In the parallel (crossed) polarization geometries we employed, we observed 5 A$_{\rm g}$ (6 B$_{\rm 1g}$) modes as shown in Fig.~\ref{fig:GR:DAC} at ambient pressure. To identify artifacts from the pressure cell setup, the measurements were made without and with the DAC at ambient pressure. We used Fano profiles for the fit (black lines), describing the data well as shown in Fig.~\ref{fig:GR:DAC}. The peak positions extracted from these two data sets collected at 0~GPa were nearly identical [also see overlapping empty and solid symbols in Figs.~\ref{fig:GR}e-h) at 0~GPa]. The signal from the samples within the DAC became weaker due to the presence of the cell and the use of a lens with a smaller magnification (reduced from 50x to 20x) as seen from the comparison of the two data sets measured at 0~GPa in Figs.~\ref{fig:GR:DAC}a-d).

When the pressure was increased to 2.4 GPa, no significant change in the Raman data was found except the hardening of the phonon frequencies (due to an increased effective spring constant between atoms by pressure). The number of peaks remained identical, confirming that the crystal structure remains unchanged up to this pressure.

Figure~\ref{fig:GR} shows Raman spectra as a function of pressure from 0 to 7.63~GPa with different geometries and polarizations. With a gradual increase of the pressure up to 3~GPa, almost all Raman phonon peaks hardened, except the A$_{\rm g}$(1) peak which softened from ~188\cm~(at 0~GPa) to 156\cm~(at 3~GPa) as shown in Figs.~\ref{fig:GR}a-b). This particular A$_{\rm g}$(1) vibration [illustrated in Fig.~\ref{fig:Phonon:calc}c] became unstable as the material approached the critical pressure around 4.1 GPa (as will be discussed in Sec.~\ref{sec:hpRaman:phonon} in more detail). Thus, this particular mode can be taken as an indicator of the structural instability with pressure.

In Fig.~\ref{fig:GR}, from 4.53~GPa and upward, all phonon peaks broaden abruptly and then split into separate peaks at higher pressures, accompanying the appearance of new phonon peaks. The new modes are most clearly visible in the Raman spectra collected at the highest pressure presented (7.62~GPa); four clearly split peaks at about 150\cm~and two separate peaks at about 550\cm~ are seen in Fig.~\ref{fig:GR}a).

The spectral ranges most strongly affected by the structural transition are highlighted as blue and green shaded areas Figs.~\ref{fig:GR}a-d). For a quantitative analysis, the spectra were fitted by Fano profiles and the fitted peak positions were marked by triangular (vertical bar) symbols before (after) the transition in Figs.~\ref{fig:GR}a-d). The evolution of the peak positions as a function of pressure for four different geometries is plotted in Figs.~\ref{fig:GR}e-h), revealing clear peak splitting and emergence of new peaks starting above the estimated critical pressure of 4.1~GPa (vertically dotted red lines).

To illustrate quantitatively the evolution of the lattice dynamics upon pressure, in Fig.~\ref{fig:GR}i) we plotted the intensities of peaks around 550\cm~[in green shaded areas in Figs.~\ref{fig:GR}c-d)] as a function of pressure. Results from two crossed polarization data sets [$\bar{\bm c}(\bm a \bm b){\bm c}$, $\bar{\bm c}(\bm b \bm a){\bm c}$] are shown using two methods: summed intensities of peaks from fits between 500~$\sim$~600\cm~and integrated intensities between 550~$\sim$~600\cm~illustrated as empty and filled symbols, respectively in Fig.~\ref{fig:GR}i). The result demonstrates a kink at about 4.1~GPa, which constitutes Raman evidence for the first-order structural transition. This critical pressure for the transition is consistent with high-pressure x-ray~\cite{Veiga2017} and neutron diffraction measurements.~\cite{Takayama2019} With the Raman data, we chose the higher wave-number region [green shaded areas in Figs.~\ref{fig:GR}c-d)) to extract the critical pressure because this spectral range has weaker phonons at ambient pressure than those in the lower wave-number region [blue shaded areas in Figs.~\ref{fig:GR}c-d)), so that the change of the Raman spectra with pressure was most clearly captured.

\begin{figure}
\begin{center}
\includegraphics[width=\linewidth]{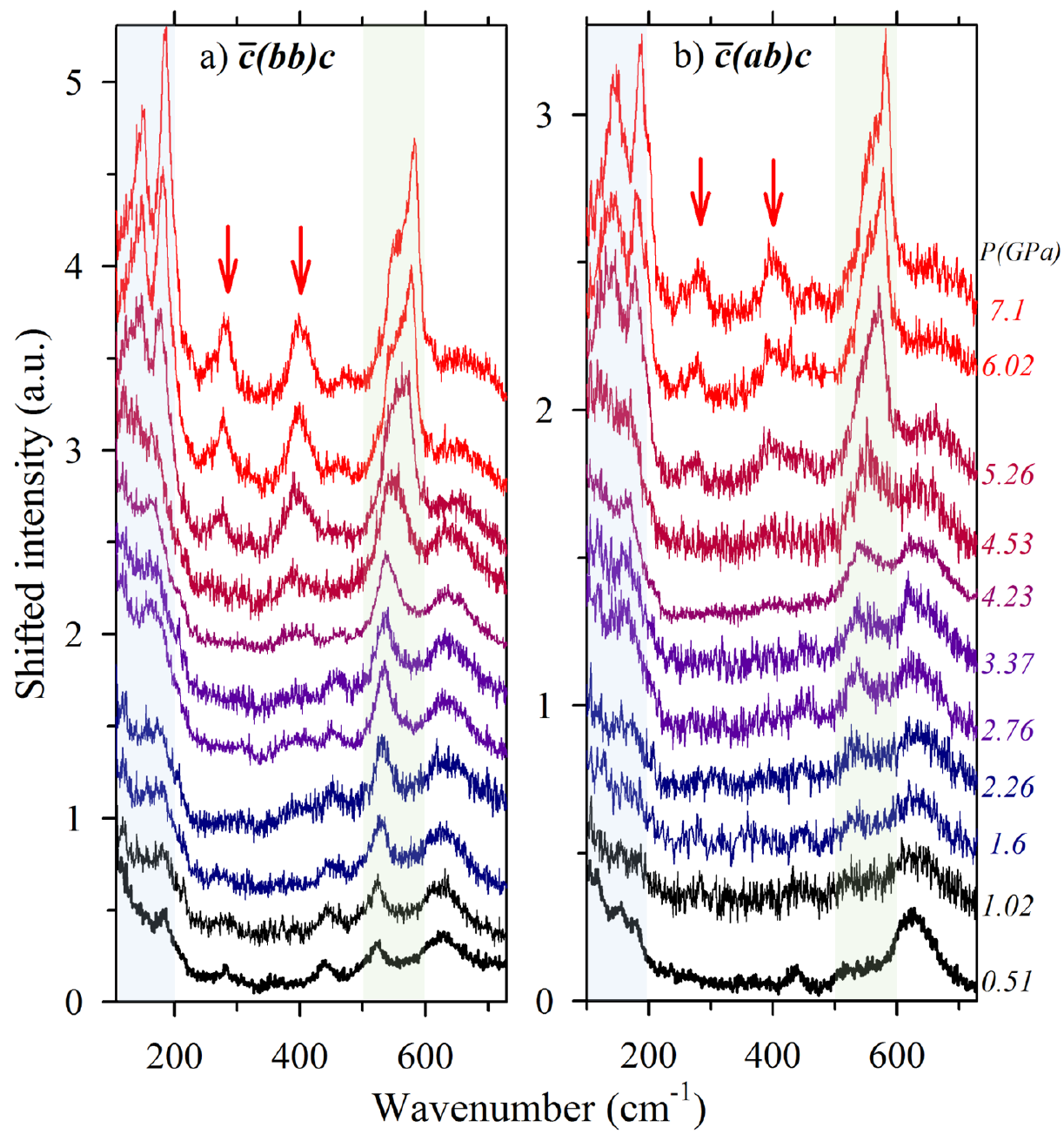}
\end{center}
\caption {Polarized Raman spectra with pressure with the red laser in a) $\bar{\bm c}(\bm b \bm b){\bm c}$ and b) $\bar{\bm c}(\bm a \bm b){\bm c}$ experimental geometry on a different crystal. Transparent blue and green shaded areas indicate the energy windows, where the spectra changed significantly with pressure. The geometry of all crossed polarizations is $\bar{\bm c}(\bm a \bm b){\bm c}$, except for the 0.51 GPa data where $\bar{\bm c}(\bm b \bm a){\bm c}$ was used. Measured pressures are noted in the right column next to the figure in GPa units. Note that both parallel and crossed polarization data probe phonons in A$_{g}$ symmetry at 4.23~GPa and above due to the Raman selection rules of the monoclinic structure at high pressures. Two red downward arrows emphasize phonons at about 282 and 402\cm, which are only weakly observed with the green laser in parallel polarization [Fig.~\ref{fig:GR}a)].
}
\label{fig:RR:pol}
\end{figure}

To further confirm the pressure-induced Raman spectra, we also used a Raman setup with the red laser. Figure~\ref{fig:RR:pol} presents the red Raman data with two polarizations, $\bar{\bm c}(\bm b \bm b){\bm c}$ and $\bar{\bm c}(\bm a \bm b){\bm c}$. In this measurement, a finer step of pressure was used with a short measurement time, revealing similar appearances of new Raman modes at high pressure. We observed the most pronounced changes in the Raman spectra for a similar range of wave numbers, highlighted as blue and green shaded areas in Figs.~\ref{fig:RR:pol}a-b), analogous to the green data in Fig.~\ref{fig:GR}a-d). The red Raman data is, thus, fully consistent with the green Raman data with a similar transition pressure. We point out that we better confirmed two Raman-active phonons at about 282\cm~and 402\cm~(marked by red downward arrows in Fig.~\ref{fig:RR:pol}), which were only weakly observed with the green laser shown in Figs.~\ref{fig:GR}a-d).

\begin{figure*}[t]
\includegraphics[width=\linewidth]{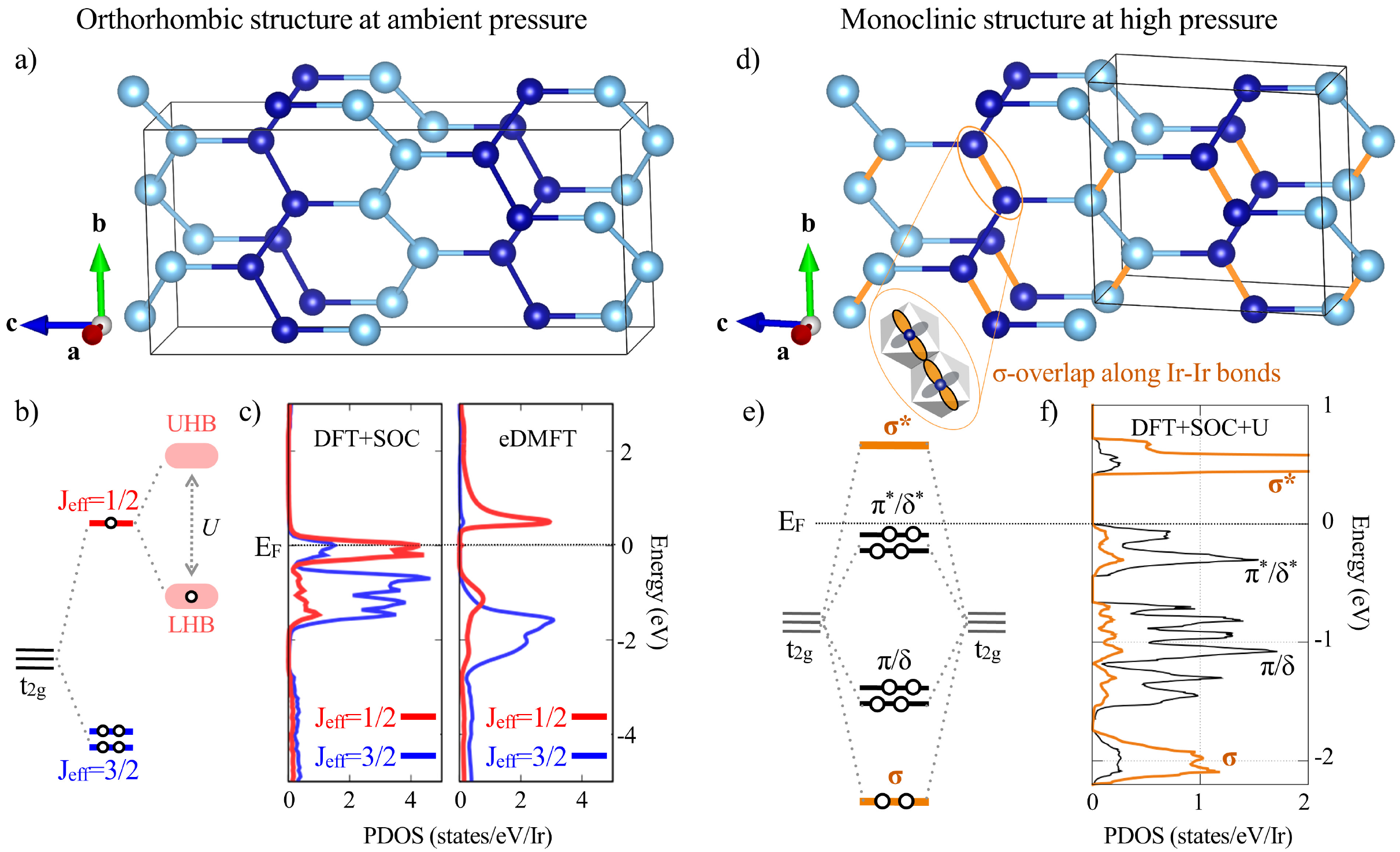}
\caption {a) Crystal structure of the relaxed orthorhombic $\beta$-Li$_2$IrO$_3$ phase at ambient pressure, where Ir sites forming nonparallel zigzag chains are depicted as dark and bright blue spheres (a relaxed DFT+SOC+$U$ structure given in Table~\ref{tableA:str:ortho} in Ref.\ \onlinecite{SM}). The black solid box is the orthorhombic unit cell. b) A schematic energy diagram of the splitting between the $J_{\rm eff}$ = 1/2 and 3/2 states and the splitting of the $J_{\rm eff}$ = 1/2 state into upper and lower Hubbard bands (UHB and LHB) by the Coulomb interaction $U$. c) Comparison between the DFT+SOC (left panel, neither $U$ nor magnetism implemented) and eDMFT (right panel) PDOS in the nonmagnetic phase. $J_{\rm eff}$ = 1/2 and 3/2 states are depicted in red and blue, respectively. The opening of Hubbard gap and the enhancement of $J_{\rm eff}$ = 1/2 - 3/2 splitting is only seen in the eDMFT result. d) The distorted hyperhoneycomb lattice of Ir ions at high pressure in the optimized crystal structure in the calculation (a relaxed DFT+SOC+$U$ structure given in Table~\ref{tableA:str:mono} in Ref.\ \onlinecite{SM}), emphasizing the dimerized Ir bonds (thick solid orange lines). The black solid box is the monoclinic unit cell. Dark and bright blue balls and sticks indicate the structural connectivity between zigzag chains. The inset shows the $\sigma$-overlapping t$_{2g}$ orbitals driving the dimer formation. e) A schematic energy diagram representing the splitting of the Ir t$_{2g}$ subspace and formation of the dimer bonding-antibonding orbitals. f) PDOS for Ir t$_{2g}$ orbitals from the DFT+SOC+$U$ result, showing the same energy level splitting given in e). The horizontal dotted line shows the Fermi energy, E$_{\rm F}$. Lithium and oxygen ions are not visualized for simplicity in a) and d).
}
\label{fig:calc}
\end{figure*}

\subsection{{\it Ab initio} calculations}
\label{sec:abinitio}
We now discuss our {\it ab initio} DFT and DMFT calculation results on lattice structures and electronic properties with and without pressure. We first address DFT+SOC+$U$ results on the phonon frequencies. To obtain accurate results, careful optimization of crystal structures in both orthorhombic and monoclinic symmetries (representing ambient and high-pressure conditions, respectively) is crucial. All the lattice parameters (i.e., magnitudes of the Bravais lattice vectors and angles between them) and internal atomic coordinates were allowed to relax [see Figs.~\ref{fig:calc}a,d)).

For the high-pressure conditions, the experimental monoclinic cell parameters~\cite{Veiga2017} were first adopted as a trial structure, and then the unit cell shape and internal ionic coordinates (with the fixed cell volume) were optimized. No symmetry conditions were enforced during the optimizations. Our results reproduced the experimental lattice parameters and internal atomic coordinates reasonably well [see Table~\ref{tableA:str:ortho} and Table~\ref{tableA:str:mono} in Supplemental Material~\cite{SM} for the full details]. It should be mentioned that the preconditioning is important to obtain reasonable crystal structures.

At ambient pressure, DFT+SOC+$U$ calculations demonstrate the essential role of SOC and the on-site Coulomb interaction in maintaining the orthorhombic close-to-ideal hyperhoneycomb structure; the orthorhombic structure can be stabilized only when the Coulomb interaction and SOC are both incorporated ($U$ = 2 eV for entire calculations) to obtain the magnetic and insulating phase.~\cite{Kim2016}

On the other hand, at high pressure, we obtain practically identical monoclinic structures with nonmagnetic (i.e., no local magnetic moments) dimerized Ir pairs regardless of the presence of SOC or $U$. This naturally implies an essential role of SOC and $U$ for the orthorhombic crystal structure at ambient pressure, and in contrast its irrelevance in the high-pressure monoclinic structure. After the optimization, the phonon energies were obtained by diagonalizing the dynamical matrix.

We now turn to calculations of the electronic structure. To gain additional insight into the electronic structures of the low- and high-pressure phases, we have performed a paramagnetic (PM) eDMFT calculation ($T$ = 232 K, $U$ = 5.0 eV, and $J$ = 0.8 eV for Ir $t_{\rm 2g}$ orbitals) to stabilize the PM Mott insulating phase, which consists of disordered and localized magnetic moments instead of noninteracting bands in DFT-based calculations. The right panel of Fig.~\ref{fig:calc}c) presents the $J_{\rm eff}$-projected density of states (PDOS) from this eDMFT calculation, revealing an evident $J_{\rm eff}$ = 1/2 character. A clear separation between the $J_{\rm eff}$ = 1/2 and 3/2 states with the gap opening can be seen, showing that the enhancement of the $J_{\rm eff}$ = 1/2 - 3/2 splitting by electron correlations is significant even in the PM phase. This is shown in the schematic energy level diagram of Fig.~\ref{fig:calc}b). It is worth mentioning that such noticeable enhancement of SOC has not been observed in previous nonmagnetic DFT+SOC results [see the left panel of Fig.~\ref{fig:calc}c)].~\cite{Kim2014,Kim2015,Kim2016}

At high pressure, we found that DFT+SOC+$U$ and DFT (without SOC and $U$) gave practically the same density of states, indicating that correlation effects become less important in the dimerized monoclinic structure. The optimized monoclinic crystal structure, starting from the experimentally determined monoclinic structure~\cite{Veiga2017}, is visualized in Fig.~\ref{fig:calc}d), where the bond lengths for the short (thick orange lines) and long Ir-Ir bonds (thin dark and light blue lines) are 2.60 and 3.05~\AA. In Fig.~\ref{fig:calc}d), we also depicted the black solid box for the monoclinic unit cell at high pressure, where one can compare with the orthorhombic unit cell at ambient pressure in Fig.~\ref{fig:calc}a) by finding how the Ir dimers are formed at high pressure using a circled inset in Fig.~\ref{fig:calc}d). This strong Ir dimerization found in the DFT+SOC+$U$ optimized structure is consistent with the experimental observation using x-rays~\cite{Veiga2017}; 2.662 and 3.012~\AA~for the short and long Ir-Ir bonds, respectively. This imposes a large ligand field on the Ir $t_{\rm 2g}$ orbitals.

Figure~\ref{fig:calc}e) sketches the energy level splitting within the Ir $t_{\rm 2g}$ dimer, where the DFT+SOC+$U$ calculated PDOS is shown in Fig.~\ref{fig:calc}f). The results clearly demonstrate an energy gap associated with a strong bonding-antibonding splitting within the $t_{\rm 2g}$ states, rendering the SOC ineffective and converting monoclinic \bLIO~into a non-magnetic band insulator. These results are consistent with a recent resonant inelastic x-ray scattering study that also indicates the pressure-induced breakdown of the spin-orbit Mott insulating state in \bLIO~\cite{Takayama2019}, and with the results of DFT+SOC+$U$ calculations.~\cite{Antonov2018}

\begin{figure}[t]
\includegraphics[width=\linewidth]{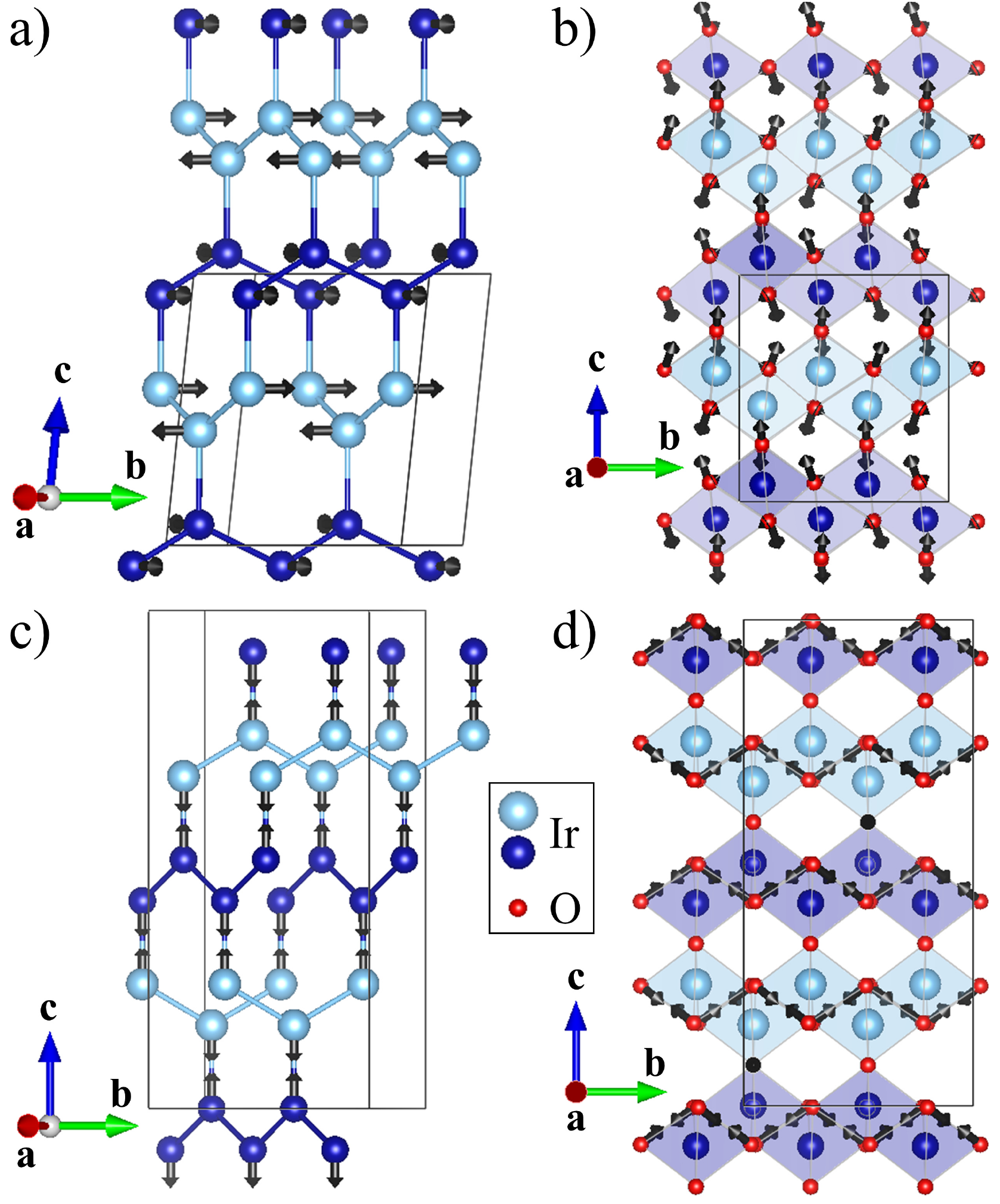}
\caption {Comparison between the eigenvectors of representative phonons at both ambient and high pressure. Calculated Raman-active A$_{\rm g}$ vibrations at high pressure in a) and b) and at ambient pressure in c) and d) are compared in the spectral ranges most affected by the structural transition. Two sets of calculated Raman modes are visualized and compared. A lower-energy Raman-active vibration is illustrated in a) 152\cm~at high pressure, which is compared with c) 194\cm~at ambient pressure. Similarly, a higher-energy Raman vibration is shown in b) 579\cm~at high pressure, which is compared with d) 587\cm~at ambient pressure. Smaller red balls are for oxygen ions.
}
\label{fig:Phonon:calc}
\end{figure}

\begin{table*}[htp]
\caption[] {Calculated Raman-active phonons by {\sc vasp} for ambient orthorhombic and high-pressure monoclinic crystal structure with a unit of\cm.}
\label{table:calc}
\begin{center}
\centering
\begin{tabular}
[c]{c c c c c c c c c c}\hline\hline  & A$_{\rm g}$(1) & A$_{\rm g}$(2) & A$_{\rm g}$(3) & A$_{\rm g}$(4) & A$_{\rm g}$(5) & A$_{\rm g}$(6) & A$_{\rm g}$(7) \\
\hline
A$_{\rm g}$ (F$ddd$)            ~&194~&281~&326~&342~&503~&516~&587~\\
\hline\hline
\\
\hline\hline
                                          ~&B$_{\rm 1g}$(1) & B$_{\rm 1g}$(2) & B$_{\rm 1g}$(3) & B$_{\rm 1g}$(4) & B$_{\rm 1g}$(5) & B$_{\rm 1g}$(6) & B$_{\rm 1g}$(7) & B$_{\rm 1g}$(8) &\\
\hline
B$_{\rm 1g}$ (F$ddd$)         ~&141~&258~&274~&332~&359~&500~&550~&604~&\\
\hline\hline
\\
\hline\hline
                                         ~&A$_{\rm g}$(1)    & A$_{\rm g}$(2) & A$_{\rm g}$(3) & A$_{\rm g}$(4) & A$_{\rm g}$(5) & A$_{\rm g}$(6) & A$_{\rm g}$(7) & A$_{\rm g}$(8) & A$_{\rm g}$(9) \\
\hline
A$_{\rm g}$ (C$2/c$)           ~&152~&215~&251~&255~&278~&314~&344~&360~&388~\\
\hline
                                         ~&A$_{\rm g}$(10) & A$_{\rm g}$(11) & A$_{\rm g}$(12) & A$_{\rm g}$(13) & A$_{\rm g}$(14) & A$_{\rm g}$(15) & A$_{\rm g}$(16) & A$_{\rm g}$(17) & A$_{\rm g}$(18)\\
\hline
                                         ~&395~&470~&477~&519~&534~&571~&579~&605~&701~\\
\hline\hline
\end{tabular}
\end{center}
\end{table*}

\subsection{Comparison between experimental and computational data}
\label{sec:hpRaman:phonon}
We are now in a position to compare our experimental and theoretical data. At ambient pressure, we experimentally observe nearly all predicted phonons, and the frequencies agree reasonably well with the calculations (see Fig.~\ref{fig:GR:DAC} and Tables~\ref{table:calc} and ~\ref{table:PP} for comparison).~\cite{SM_calc} On the other hand, a similar comparison is not possible in the high pressure phase due to the low symmetry of the lattice and the large number of phonon modes observed. However, the measured Raman spectra can be qualitatively understood by relating the data from ambient to high pressure to the underlying crystal structures.

Specifically, we compared experimental and computational data of representative phonons in the highlighted regions of wave numbers at around 150 and 550\cm~in Fig.~\ref{fig:GR}a). At ambient pressure, we examined two Raman-active phonons illustrated in Fig.~\ref{fig:Phonon:calc}c) and d) with calculated frequencies of 194\cm~and 587\cm~ (see Table~\ref{table:calc} for a list of Raman-active phonons calculated for comparison with the experimental list in Table~\ref{table:PP}). The eigenvectors are dominated by Ir vibrations at the lower frequency (due to heavier masses) and oxygen vibrations at the higher frequency (due to lighter masses).

At high pressure, the strong dimerization of Ir bonds and the corresponding distortion of oxygen octahedral cages [as visualized in Fig.~\ref{fig:calc}d)] are expected to greatly affect the phonon modes. Specifically, the dimerized bond of Ir ions (the heaviest ions) is anticipated to alter mostly the low-energy Raman spectra, and indeed we observed a significant modification of the phonon spectra between 100~$\sim$~200\cm. This is confirmed by the Ir-dominant atomic vibrations (with negligible amount of Li and O vibrations) found in the DFT+SOC+$U$ calculations: see $A_{\rm g}$-phonon modes compared in Fig.~\ref{fig:Phonon:calc}c) (at 194\cm~from F$ddd$) and \ref{fig:Phonon:calc}a) (at 152\cm~from C$2/c$) for ambient and high pressure, respectively. Also, a similarly sudden change of the phonon spectra at high pressure was experimentally observed at higher energy, which should be naturally linked to the lighter oxygen ions. Indeed, this can be verified from the DFT phonon modes shown in Fig.~\ref{fig:Phonon:calc}d) (at 587\cm~from F$ddd$, the highest A$_{\rm g}$ mode calculated) and ~\ref{fig:Phonon:calc}b) (at 579\cm~from C$2/c$).

We should note that these vibrations are qualitatively distinct even though they have similar frequencies: their eigenvectors are almost perpendicular to each other [compare Fig.~\ref{fig:Phonon:calc}a) and c), as well as Fig.~\ref{fig:Phonon:calc}b) and d)]. We confirmed that two phonon modes calculated at high pressure [Fig.~\ref{fig:Phonon:calc}a) and Fig.~\ref{fig:Phonon:calc}b)] are not present among the calculated Raman modes at ambient pressure. That is, both the Raman experiments and the phonon calculations reveal the distinct nature of the atomic motions in ambient and high-pressure structures.

Our results fit into a conceptual framework that attributes a variety of related phenomena in 4$d$- and 5$d$-electron materials to a competition between intermetallic covalency and magnetism.~\cite{Streltsov2016} Within this scheme, the formation of dimerized bonds can be understood as a consequence of the reduced kinetic energy of the electrons within the dimer at the expense of the formation of localized magnetic moments. Our observations in \bLIO~under pressure are consistent with the notion, so that the shrinking of Ir~-~Ir distances with pressure sharply increases the hopping between $d$~-~$d$ orbitals, driving a first-order structural transition. This theory~\cite{Streltsov2016} has also been applied to 3$d$ transition metal compounds such as CrO$_{2}$,~\cite{SKim2012} where a dimerized monoclinic structure is theoretically predicted at about 70~GPa.~\cite{Streltsov2016,SKim2012} The lower critical pressure in \bLIO~may be due to the much more extended 5$d$ orbitals with the larger $d$~-~$d$ hopping compared to the 3$d$ example. These considerations can be generalized to the family of \LIO~based on similar observations. For example, high-pressure resonant inelastic x-ray scattering experiments on \aLIO~\cite{Clancy2018} have found the breakdown of the $J_{\rm eff}=1/2$ picture between 0.1~GPa and 2~GPa, followed by a structural transition to the dimerized ground state at above 3~GPa.

\section{Summary}
\label{sec:summary}
We performed a combined analysis using high-pressure Raman scattering and {\it ab initio} calculations on the hyperhoneycomb iridate \bLIO. Using Raman scattering under pressure, we experimentally observed the broadening and splitting of phonon peaks and the appearance of new modes at high pressure, explained by a symmetry lowering via a first-order structural transition. This is further confirmed by phonon calculations comparing the lattice dynamics at both pressures. The calculations clearly demonstrated the breakdown of the $J_{\rm eff}$ = 1/2 state due to the Ir-Ir bond dimerization that leads to the high-pressure monoclinic phase. This observation can be interpreted in terms of a competition between intermetallic covalency and the formation of Ir local moments. Our results demonstrate that Raman scattering is an effective probe of pressure-induced structural and electronic phase transitions in materials with 4$d$ and 5$d$ valence electrons.

\section{Acknowledgments}
We appreciate technical support from J. Nuss for x-ray characterization to cross-check the orientation of crystals, from U. Engelhardt for preparing high-pressure gaskets, and especially from K. Syassen for providing the DAC and helping to mount crystals as well as from J. Na for help with the Raman measurements. H.-S. K. thanks C. Dreyer for providing his script on phonon mode projections. We acknowledge funding from the Deutsche Forschungsgemeinschaft (DFG, German Research Foundation) – Projektnummer 107745057 – TRR 80. H.-S. K., K. H., and D. V. are supported by NSF DMREF grant DMR-1629059.

\begin{table*}
\caption[] {Peak positions extracted from the fit using the green Raman data collected upon pressure for four polarizations as shown in Fig.~\ref{fig:GR}. The unit of frequencies is\cm.}
\label{table:PP}
\begin{center}
\centering
\begin{tabular}
[c]{c c | c c c c c c c c c c c c c c}\hline\hline
$\bar{\bm c}(\bm a \bm a){\bm c}$~&P~(GPa)~&-~&-~&A$_{\rm g}$(1)~&-~&-~&-~&A$_{\rm g}$(2)~&-~&-~&-~&-~&A$_{\rm g}$(4)~&A$_{\rm g}$(6)~&A$_{\rm g}$(7)~\\
\hline
~&0~&-~&-~&188~&-~&-~&-~&330~&-~&-~&-~&-~&518~&596~&661~\\
~&0~&-~&-~&188~&-~&-~&-~&330~&-~&-~&-~&-~&520~&596~&663~\\
~&2.4~&-~&-~&159~&-~&-~&-~&-~&-~&-~&-~&-~&533~&606~&688~\\
~&3~&-~&-~&156~&-~&-~&-~&-~&-~&-~&-~&-~&536~&619~&668~\\
\hline
~&P$>$P$_{c}$~&A$^{*}_{g}$(1)~&A$^{*}_{g}$(2)~&A$^{*}_{g}$(3)~&A$^{*}_{g}$(4)~&A$^{*}_{g}$(5)~&A$^{*}_{g}$(6)~&A$^{*}_{g}$(7)~&A$^{*}_{g}$(8)~&A$^{*}_{g}$(9)~&A$^{*}_{g}$(10)~&A$^{*}_{g}$(11)~&A$^{*}_{g}$(12)~&A$^{*}_{g}$(13)~&A$^{*}_{g}$(14)~\\
\hline
~&4.53~&143~&164~&189~&-~&-~&-~&-~&392~&434~&536~&566~&-~&635~&672~\\
~&5~&143~&164~&189~&-~&-~&-~&-~&392~&434~&536~&566~&601~&635~&672~\\
~&6.3~&144~&163~&184~&200~&-~&-~&-~&401~&455~&556~&579~&607~&673~&687~\\
~&7.62~&149~&163~&189~&203~&-~&-~&-~&405~&467~&562~&585~&613~&687~&697~\\
\hline\hline
$\bar{\bm c}(\bm b \bm b){\bm c}$~&P~(GPa)~&-~&-~&A$_{\rm g}$(1)~&-~&-~&-~&A$_{\rm g}$(2)~&-~&-~&-~&-~&A$_{\rm g}$(4)~&A$_{\rm g}$(6)~&A$_{\rm g}$(7)~\\
\hline
~&0~&-~&-~&187~&-~&-~&-~&357~&-~&-~&-~&-~&519~&596~&665~\\
~&0~&-~&-~&183~&-~&-~&-~&357~&-~&-~&-~&-~&520~&598~&672~\\
~&2.4~&-~&-~&148~&-~&-~&-~&356~&-~&-~&-~&-~&530~&621~&667~\\
~&3~&-~&-~&150~&-~&-~&-~&375~&-~&-~&-~&-~&533~&630~&670~\\
\hline
~&P$>$P$_{c}$~&A$^{*}_{g}$(1)~&A$^{*}_{g}$(2)~&A$^{*}_{g}$(3)~&A$^{*}_{g}$(4)~&A$^{*}_{g}$(5)~&A$^{*}_{g}$(6)~&A$^{*}_{g}$(7)~&A$^{*}_{g}$(8)~&A$^{*}_{g}$(9)~&A$^{*}_{g}$(10)~&A$^{*}_{g}$(11)~&A$^{*}_{g}$(12)~&A$^{*}_{g}$(13)~&A$^{*}_{g}$(14)~\\
\hline
~&4.53~&118~&139~&164~&184~&-~&-~&-~&388~&434~&547~&561~&-~&637~&669~\\
~&5~&135~&159~&179~&196~&-~&-~&-~&394~&443~&552~&573~&-~&642~&675~\\
~&6.3~&143~&163~&187~&201~&-~&280~&-~&401~&463~&558~&582~&-~&645~&681~\\
~&7.62~&150~&168~&195~&205~&-~&284~&-~&405~&-~&565~&591~&-~&-~&-~\\
\hline\hline
$\bar{\bm c}(\bm a \bm b){\bm c}$~&P~(GPa)~&-~&-~&B$_{\rm 1g}$(1)~&-~&B$_{\rm 1g}$(4)~&-~&B$_{\rm 1g}$(5)~&-~&B$_{\rm 1g}$(6)~&-~&-~&B$_{\rm 1g}$(7)~&-~&B$_{\rm 1g}$(8)~\\
\hline
~&0~&-~&-~&153~&-~&-~&-~&380~&-~&522~&-~&-~&559~&-~&625~\\
~&0~&-~&-~&151~&-~&319~&-~&360~&-~&520~&-~&-~&569~&-~&627~\\
~&2.4~&-~&-~&157~&-~&-~&-~&-~&-~&534~&-~&-~&609~&-~&677~\\
~&3~&-~&-~&157~&-~&-~&-~&-~&-~&539~&-~&-~&613~&-~&685~\\
\hline
~&P$>$P$_{c}$~&A$^{*}_{g}$(1)~&A$^{*}_{g}$(2)~&A$^{*}_{g}$(3)~&A$^{*}_{g}$(4)~&A$^{*}_{g}$(5)~&A$^{*}_{g}$(6)~&A$^{*}_{g}$(7)~&A$^{*}_{g}$(8)~&A$^{*}_{g}$(9)~&A$^{*}_{g}$(10)~&A$^{*}_{g}$(11)~&A$^{*}_{g}$(12)~&A$^{*}_{g}$(13)~&A$^{*}_{g}$(14)~\\
\hline
~&4.53~&122~&155~&165~&180~&-~&-~&-~&394~&-~&535~&560~&607~&624~&674~\\
~&5~&127~&147~&178~&195~&-~&278~&-~&398~&-~&555~&572~&602~&645~&681~\\
~&6.3~&130~&149~&187~&200~&-~&282~&-~&403~&-~&564~&583~&611~&656~&694~\\
~&7.62~&149~&166~&187~&200~&-~&282~&-~&403~&-~&566~&586~&613~&661~&698~\\
\hline\hline
$\bar{\bm c}(\bm b \bm a){\bm c}$~&P~(GPa)~&-~&-~&B$_{\rm 1g}$(1)~&-~&B$_{\rm 1g}$(4)~&-~&B$_{\rm 1g}$(5)~&-~&B$_{\rm 1g}$(6)~&-~&-~&B$_{\rm 1g}$(7)~&-~&B$_{\rm 1g}$(8)~\\
\hline
~&0~&-~&-~&152~&-~&323~&-~&367~&-~&501~&-~&-~&557~&-~&629~\\
~&0~&-~&-~&152~&-~&314~&-~&363~&-~&500~&-~&-~&561~&-~&630~\\
~&2.4~&-~&-~&160~&-~&-~&-~&-~&-~&533~&-~&-~&607~&-~&672~\\
~&3~&-~&-~&160~&-~&-~&-~&-~&-~&533~&-~&-~&609~&-~&686~\\
\hline
~&P$>$P$_{c}$~&A$^{*}_{g}$(1)~&A$^{*}_{g}$(2)~&A$^{*}_{g}$(3)~&A$^{*}_{g}$(4)~&A$^{*}_{g}$(5)~&A$^{*}_{g}$(6)~&A$^{*}_{g}$(7)~&A$^{*}_{g}$(8)~&A$^{*}_{g}$(9)~&A$^{*}_{g}$(10)~&A$^{*}_{g}$(11)~&A$^{*}_{g}$(12)~&A$^{*}_{g}$(13)~&A$^{*}_{g}$(14)~\\
\hline
~&4.53~&130~&144~&165~&182~&-~&-~&-~&392~&-~&532~&563~&604~&637~&676~\\
~&5~&139~&158~&177~&189~&-~&279~&-~&394~&-~&553~&570~&603~&643~&681~\\
~&6.3~&150~&161~&184~&200~&-~&279~&-~&401~&-~&563~&579~&609~&663~&693~\\
~&7.62~&152~&165~&188~&203~&-~&283~&-~&404~&-~&567~&584~&614~&660~&697~\\
\hline\hline
\end{tabular}
\end{center}
\end{table*}

\vspace{1cm}

\textbf{ \em{Supplemental Material for}} \textbf {Lattice dynamics and structural transition of the hyperhoneycomb iridate \bLIO investigated by high-pressure Raman scattering}

\vspace{2 mm}

This Supplemental Material present the full technical details of single crystal growth and characterization in Section S1, Raman measurements to characterize beam-heating in Section S2, Raman measurements in Section S3, followed by details of structural relaxations in Section S4 and calculations of phonon frequencies in Section S5, and computational details in Section S6. Tables provide relative fitted peak positions from the beam-heating measurements and DFT/DMFT-optimized crystal structures at both ambient and high pressure.

\maketitle
\vspace{2 mm}

\section*{S1. Single crystal growth and characterization}
\label{sm:Xtals}
\vspace{-0.2cm}

High-quality single crystals were synthesized by a flux-method explained elsewhere.~\cite{Takayama:beta:Li213} The crystallographic axes of the single crystals were determined by single crystal x-ray diffraction and polarization-resolved Raman measurements by means of the Raman selection rules.~\cite{Glamazda2016} The crystal morphology was plate-like with the $\bm{c}$-axis perpendicular to the plane, similar to its structural polytype~\gLIO.~\cite{Modic2014} [see Fig.~\ref{sm:fig:sample}a) for a microscopic image]. Typical crystal dimensions used for Raman experiments are about 30$\mu$m $\times$ 20$\mu$m $\times$ 15$\mu$m.

\section*{S2. Raman measurements to characterize beam-heating}
\label{sm:BH}
\vspace{-0.2cm}

To estimate extrinsic shifts of phonon peaks in the room temperature measurements by a local beam-heating of crystals, we carefully determined the maximised Raman signal to noise ratio that we could get while we avoided the beam-heating for all experimental setting that we used: 50x, 20x without the Diamond Anvil Cell (DAC) and 20x with the DAC in both green and red lasers. Figure~\ref{sm:fig:sample}a) shows a typical morphology of the crystal used in the experiment with the corresponding unit cell with the iridium network in Fig.~\ref{sm:fig:sample}b).

Figure~\ref{sm:fig:BH} summarizes the softening of selected phonon peaks with an elevating Raman laser power: we chose A$_{\rm g}$(1), A$_{\rm g}$(4), A$_{\rm g}$(6) (A$_{\rm g}$(4), A$_{\rm g}$(6)) with the red laser (with the green laser) as they are stronger than other peaks. By fitting normalised fitted peak positions [explicitly given in Table~\ref{table:BH:PP} from a pseudo Voigt fit, a combined function with Lorentzian and Gaussian profiles], we found that the threshold Raman power (not inducing the beam-heating effect) is about~0.7~mW without the DAC [Fig.~\ref{sm:fig:BH}a)] and about~2~mW with the DAC [Fig.~\ref{sm:fig:BH}b)]: due to a bigger attenuation of light intensity by the DAC, the threshold laser power increases with the DAC [as compared in Figs.~\ref{sm:fig:BH}a) and \ref{sm:fig:BH}b)]. To avoid the beam-heating, these threshold Raman laser powers were used for all relevant Raman measurements presented in this paper.

\renewcommand{\thefigure}{S1}
\begin{figure}[t]
\begin{center}
\includegraphics[width=\linewidth]{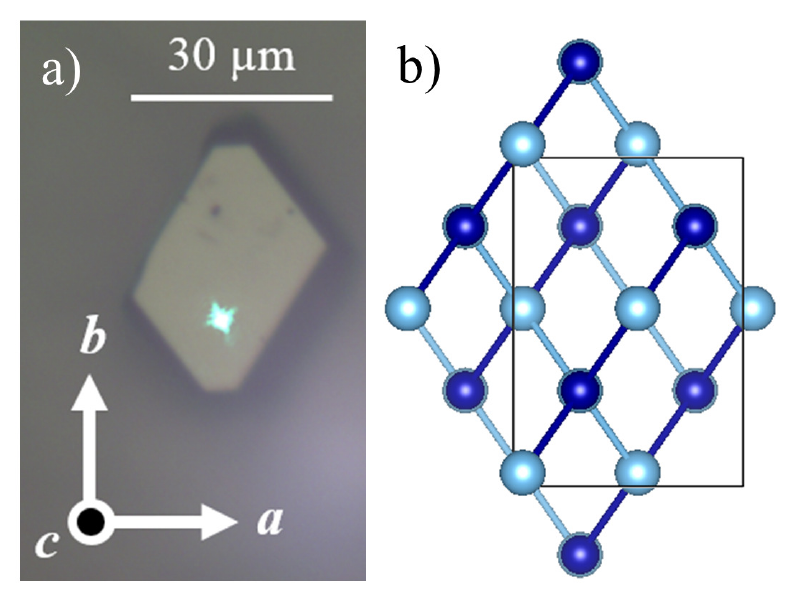}
\end{center} \caption {(color online) a) A microscopic image of a representative \bLIO crystal with a green laser at ambient pressure without the DAC, focused with a 50x microscope. b) The same orientation of the unit cell with Ir ions (blue and dark blue balls). Oxygen and lithium ions are not displayed for a simplicity.
}
\label{sm:fig:sample}
\end{figure}

\renewcommand{\thefigure}{S2}
\begin{figure}[t]
\begin{center}
\includegraphics[width=\linewidth]{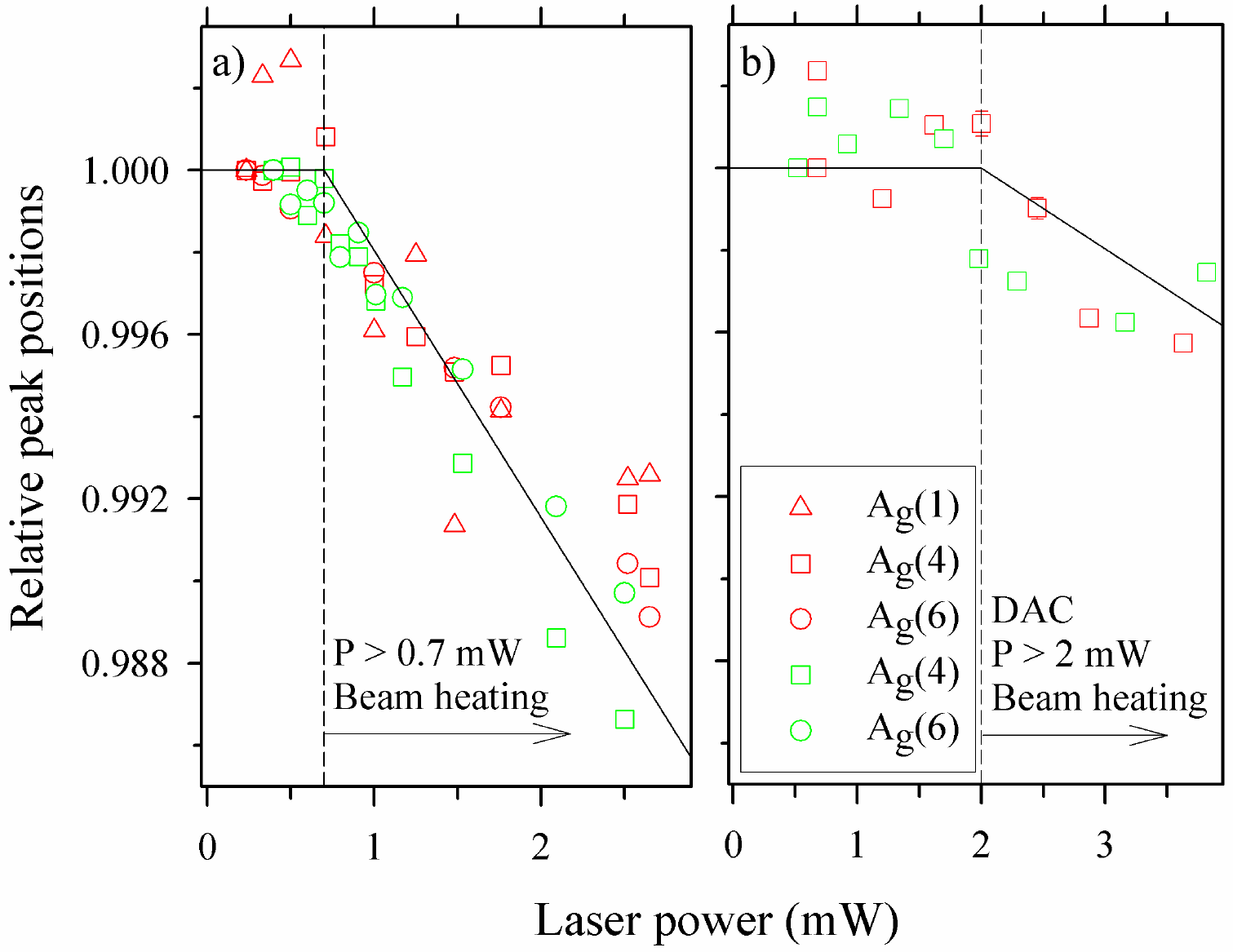}
\end{center} \caption {(color online) Normalized positions of selected Raman peaks in terms of laser powers at ambient pressure a) without the DAC and b) with the DAC. Green (red) symbols are phonon peaks obtained by using the green (red) Raman laser. Two vertically dashed black lines indicate the estimated (based on the fits with solid black lines) strongest laser power at about 0.7~mW (2~mW) without (with) the DAC, which do not induce an artificial beam-heating effect.
}
\label{sm:fig:BH}
\end{figure}

\begin{table*}
\caption[] {Fitted peak positions with the red and green Raman data upon the laser power. A lens of 20x was used with the DAC and a 50x lens was utilized without the DAC. The R (G) symbol in the first row means the red (green) laser. The unit of frequencies is\cm. The typical fitting error bars are in the two decimal places.}
\label{table:BH:PP}
\vspace{-0.2cm}
\begin{center}
\centering
\begin{tabular}
[c]{c | c c || c| c c c || c |c || c |c}\hline\hline  Power & A$_{\rm g}$(4)G & A$_{\rm g}$(6)G & Power & A$_{\rm g}$(1)R & A$_{\rm g}$(4)R & A$_{\rm g}$(6)R & Power (DAC) & A$_{\rm g}$(4)G & Power (DAC) & A$_{\rm g}$(4)R \\
\hline
0.40~mW~&518.540~&597.422~&2.65~mW~&181.899~&514.702~&617.211~&0.52~mW~&532.072~&0.675~mW ~&517.569\\
0.50~mW~&518.579~&596.928~&2.52~mW~&181.883~&515.624~&618.024~&0.68~mW~&532.866~&0.682~mW ~&518.797\\
0.60~mW~&517.970~&597.131~&1.76~mW~&182.185~&517.382~&620.397~&0.92~mW~&532.382~&1.2~mW     ~&517.181\\
0.70~mW~&518.438~&596.944~&1.48~mW~&181.674~&517.299~&620.999~&1.34~mW~&532.846~&1.62~mW   ~&518.112\\
0.80~mW~&517.614~&596.153~&1.25~mW~&182.880~&517.749~&              ~&1.7  ~mW~&532.451~&2~mW         ~&518.128\\
0.91~mW~&517.452~&596.514~&1.00~mW~&182.543~&518.406~&622.438~&1.98~mW~&530.899~&2.45~mW   ~&517.061\\
1.01~mW~&516.893~&595.611~&0.71~mW~&182.964~&520.274~&              ~&2.29~mW~&530.606~&2.87~mW    ~&515.678\\
1.17~mW~&515.929~&595.570~&0.50~mW~&183.748~&519.837~&623.417~&3.16~mW~&530.075~&3.63~mW    ~&515.365\\
1.53~mW~&514.842~&594.527~&0.33~mW~&183.679~&519.715~&623.918~&3.82~mW~&530.721~&                     ~&~\\
2.09~mW~&512.637~&592.528~&0.23~mW~&183.258~&519.852~&623.999~&4.7  ~mW~&530.383~&                     ~&~\\
2.50~mW~&511.614~&591.268~&                 ~&              ~&              ~&              ~&                  ~&             ~&                     ~&~\\
\hline\hline
\end{tabular}
\end{center}
\end{table*}

\section*{S3. Raman measurements}
\label{sm:Raman}
\vspace{-0.2cm}

Raman experiments with the green laser were performed with the 514 nm excitation line of an argon/krypton laser using a JobinYvon T64000 spectrometer with an energy resolution of~$\sim$~2.4\cm~(measured by a neon lamp). The measurements with the red laser used the 632.8 nm line of a HeNe mixed gas laser and a Labram (Horiba Jobin-Yvon) single-grating spectrometer.~\cite{Gretarsson2016, Souliou2017}

First, we checked the Raman spectra at ambient pressure with various polarizations: measurements did not show any meaningful differences in the Raman spectra, indicating homogeneous sample quality and compositions for good crystals.

We also obtained the Raman data with the rotated crystal with various in-plane angles along the perpendicular direction (along the $\bm{c}$-axis) of the $(\bm a \bm b)$-plane-oriented crystal [see Fig.~\ref{sm:fig:sample}a)]: we confirmed that small misalignment within the plane ($\Delta \theta~\lesssim$~15$^{\circ}$) did not give very noticeable change in the spectra, ensuring reliable Raman spectra collected at even high pressure: note that a negligible misorientation of the in-plane orientation ($\lesssim$~5$^{\circ}$) was unavoidable during high-pressure Raman measurements.

The diameter of the beam was typically $\lesssim$~5~$\mu$m at ambient pressure without the DAC, measured by a varying size of circles of gold deposited to the Al$_{2}$O$_{3}$~substrate using a sharp contrast of Raman signals from gold~\cite{ref:gold}(a broad continuous intensity) and Al$_{2}$O$_{3}$~\cite{Li2010} (a set of sharp phonon peaks). Without the DAC, we confirmed that the effective size of Raman light is approximately similar to the size of circular light observed under the microscope. Based on this, we estimated the beam size inside the DAC, to be $\lesssim$~30~$\mu$m (estimated only with the red Raman light as green Raman lights give a much smaller beam size even with the DAC).

Systematic and accurate measurements were pursued by controlling various experimental conditions. For instance, the identical microscopic lens were used for all measurements: Nikon 50x/0.45 Super Long Working Distance (SLWD) and 20x/0.35 SLWD lens (the largest magnification lens available to us to be compatible with our diamond anvil cell) to use intentionally the same attenuation rate of light. It was because different types of lens would have different attenuation rates for the given light, so the beam-heating rate, which was obtained from the analysis shown in Fig.~\ref{sm:fig:BH}, would be modified accordingly.

With the green Raman light (not necessarily with the red Raman light), the continuous flow of Ar-gas has been implemented to effectively suppress Raman signals from the vibrational air scattering, mostly below about 150\cm, which was crucial to reliably identify and trace phonon peaks at the low-energy transfer especially at high pressures. In green and red Raman experiments, different single crystals were used with a similar sample quality.

Moreover, Raman measurements on other beam positions at both ambient and finite pressures with the DAC were tested, finding only a mere change in the background signal. Small linear background signals (mostly coming from the DAC) were subtracted for some high-pressure data when necessary for a better representation. A small variation of the background signal at different pressure and polarization seems to be originated from a slight redistribution of the medium liquid (see Section S2B for details) when the new pressure was applied and/or the shape and size of the Raman light was changed depending on the incident polarization of the light. All Raman measurements were made with a high-resolution (1800~grooves/mm) setting to measure the Raman spectra more precisely.

\subsection*{A. Polarized measurements with green laser ~\\(514.5 nm)}
\label{sm:Raman:green}
We should mention that there are some ambiguities to identify weak and overlapped Raman peaks at high pressures. For example, a lesser number of Raman peaks has been experimentally measured reliably, compared with 18 Raman-active phonon modes obtained from the calculation [compare Table I and II in the main text] possibly due to their weak Raman signals. In fact, we observed very weak peak-like and shoulder features in the Raman spectra at higher pressures, but their tiny intensities for the whole pressure range explored did not allow us to do a reliable fit, so they were not marked in Fig. 2 and Table II in the main text. Moreover, in Fig. 2 in the main text, it was nearly impossible to do a reliable fit with collected Raman data at P~=~4.53 and 5~GPa due to weak intensities of new Raman peaks with an increased signal from the DAC, thus we best estimated peak positions at P~=~4.53 and 5~GPa reversely from peak positions reliably identified from P~=~6.3 and 7.62~GPa, whose peaks were much better defined. Furthermore, we cannot completely rule out a possibility of mixing of sample peaks with the DAC peaks: i.e., a peak-like signal between 220\cm~and 250\cm~for a $\bar{\bm c}(\bm b \bm b){\bm c}$ polarization in Fig. 2b) in the main text already present even at 2.4~GPa below the transition.

Asymmetric profiles of some peaks (i.e., peaks at lower wave numbers) could be also from the combination of multiple peaks nearby, or from the coupling with electronic response with the Ir $J_{\rm eff}=1/2$ local moments similarly seen in \sio.~\cite{Gretarsson2016} For testing the latter case, the fitting with a Fano asymmetry profile~\cite{Fano1961} was attempted, but did not give any noticeable trend in the fitted parameters (i.e., linewidths). If the same physics should apply to this compound, the absence of this coupling is probably due to weak Raman intensities coexistent with increased background signals when equipped with the DAC.

\subsection*{B. High-pressure measurements}
\label{sm:Raman:HP}
High-pressure Raman measurements were performed with the DAC. Diamond anvils had culet diameters of 0.4~mm and were of the ultra-low luminescence type. The stainless steel gaskets were preindented to 100~$\mu$m thickness and a hole of 175~$\mu$m diameter was drilled into each gasket by spark erosion. The hole was designed to ensure enough space for the thick \bLIO~single crystals. Several attempts with thinner gaskets failed by breaking samples at intermediate pressures during Raman measurements.

High-pressure Raman data showed weaker signals when equipped with the DAC possibly due to the strong background signal from the DAC [see Fig. 1 in the main text], an enhanced light attenuation by the DAC and a less focusing light due to a decreased magnification of available lens (from 50x to 20x): our high-pressure setup with the DAC was not compatible with the focal length of the 50x lens, whereas Raman signals without the DAC allowed a larger magnification lens (50x). The latest factor increased a typical measurement time ($\gtrapprox$~10 hours) for a single Raman data [i.e., Fig. 2 in the main text] at one pressure and polarization to maximize the signal to noise ratio.

Porto's notation~\cite{Damen1966} was utilised to describe the configuration of the Raman scattering experiment (in a backscattering geometry with the light propagating along the crystalline $\bm{c}$-axis). It expresses the orientation of the crystal with respect to the polarization of the Raman laser in both exciting and analysing directions, in a form of $k_{i}(E_{i} E_{s})k_{s}$, where $k_{i}$ ($k_{s}$) is the direction of incident (scattered) light and $E_{i}$ ($E_{s}$) is the polarization of incident (scattered) light, respectively.

At high pressure, the crystallographic axes of the monoclinic structure are different from those in ambient orthorhombic structure since the $\bm{c}$-axis is no longer parallel to the the vertical axis of the laboratory frame, but tilted by 16.777$^{\circ}$ from the normal direction.~\cite{Veiga2017} However, we did not observe any significant difference in the measured spectra for $\bar{\bm c}(\bm a \bm b){\bm c}$, $\bar{\bm c}(\bm b \bm a){\bm c}$ polarizations as shown in Fig. 2 in the main text, indicating an insensitivity of this tilted angle of the ${\bm c}$-axis in our measurements. This is consistent with our previous characterization measurements, which only showed some meaningful variations in the Raman spectra when the crystal was rotated by $\sim~$15$^{\circ}$ in the ($\bm{ab}$)-plane at ambient pressure [as explained in S3]: this makes our polarization analysis reliable even at high pressures.

By symmetry analysis, Raman tensors~\cite{BBO_server} of high-pressure monoclinic structure~\cite{Veiga2017} are given as
\begin{equation}
I_{001}({\rm A_{g}})=
\begin{pmatrix}
${\cal A}$ & ${\cal D}$        &  \\
${\cal D}$ &${\cal B}$ &  \\
         &         &${\cal C}$
\end{pmatrix} \nonumber
\end{equation}

\begin{equation}
I_{001}(B_{g})=
\begin{pmatrix}
                   &               &${\cal E}$ \\
                   &               &${\cal F}$ \\
${\cal E}$  &${\cal F}$ &
\end{pmatrix}, \label{eq:Raman}
\end{equation}
where A, B, C, D, E and F are Raman intensity components and a subscript is the direction of the propagating light (the monoclinic $\bm{c}$-axis).

\section*{S4. Structural relaxations}
\label{sm:calc:relaxations}
\vspace{-0.2cm}

At ambient pressure, both SOC and $U$ were essential (as explained in the main text) to stabilize the experimental structure. Otherwise, the calculations in the absence of either $U$ or SOC found that an initial orthorhombic structure (close to the ideal hyperhoneycomb structure) became unstable and evolved into a new type of Ir-dimerized orthorhombic crystal structure at ambient pressure, destroying the Ir $J_{\rm eff}=1/2$ local moments [dimerized along the $\bm c$-axis in Fig. 4a) in the main text]. On the other hand, when we kept the converged electronic structure with the $J_{\rm eff}=1/2$ moments and pressurized the unit cell (i.e., optimizing the cell parameters and internal coordinates with a smaller fixed volume), the orthorhombic phase without the Ir dimerization was maintained up to 10~GPa~\cite{Kim2016} as a local minima state.

\section*{S5. Calculations of phonon frequencies}
\label{sec:phon:calc}
\vspace{-0.2cm}

\subsection*{A. DFT+SOC+$U$ results}
\label{sm:phon:DMT}
In this section, we discuss our theoretical attempts to understand origins of their mismatches in calculations and experiments. Comparing in Table I and Table II in the main text, at ambient pressure, the calculated phonon frequencies agree well with the observed phonons fitted from the Raman data except for two peaks in the spectral range between 600 and 680\cm~in the A$_{\rm g}$ channel. In particular, the frequency difference between the highest measured and calculated A$_{\rm g}$ modes is about 100\cm, which is consistently reproduced by alternative DFT+SOC+$U$ calculation with {\sc wien2k} code. On the other hand, at high pressure, the calculated phonon energies agree better with the measured peaks fitted from the Raman data in Figs. 2a-d) in the main text (compare Table I and II in the main text): the agreement is slightly worse at a lower energy, possibly due to less important (but non-negligible) roles of $U$ and SOC in the dimerized structure.

We should also point out that although the overall calculating phonon frequencies match better at high pressure, a marginal mismatch of frequencies between the data and calculation is also attributed to the difference of pressures used for the comparison: a higher pressure (7.62~GPa) of the experimental data than the pressure used in the {\it ab initio} calculations at (presumably) 4~GPa.~\cite{Veiga2017}

\subsection*{B. Dynamical mean-field results on the highest $A_{\rm g}$ mode at ambient pressure}
\label{sm:phon:DMFT}
To understand the origin of the largest discrepancy between DFT+SOC+$U$ and experimental Raman data for the high energy A$_{\rm g}$ mode at ambient pressure, we also employed a method that can better describe strong correlation physics, in particular the non-perturbative nature of Mott insulator in the paramagnetic state; we present dynamical mean-field result on the highest $A_{\rm g}$ mode at ambient pressure.

For this calculation, we used the charge-self-consistent DFT+embedded dynamical mean-field theory (eDMFT) method,~\cite{eDMFT,Haule2010,Haule2018} (combined with {\sc wien2k}\cite{wien2k}) including SOC to describe the {\it paramagnetic} Mott phase of the orthorhombic structure [see S6B for computational details]. The crystal structure optimized within eDMFT (at $T$ = 232 K) also showed a reasonable agreement with the experimental and DFT+SOC+$U$ optimized structures [see Table~\ref{tableA:str:ortho} for details]. A finite displacement method for the highest $A_{\rm g}$ phonon mode was then used to draw the free energy versus the displacement curve for the calculation of the phonon frequency. As a result, the paramagnetic eDMFT predicts the frequency to be 556.6\cm.

Interestingly, this result is very close to the {\sc wien2k} magnetic DFT+$U$ result, which is 550.9\cm. This small ($\approx$ 5\cm) difference in the frequency between the paramagnetic eDMFT and magnetic DFT+SOC+U results could indicate a negligible coupling between the magnetism and the lattice. The value obtained from the {\sc vasp} method is about 587\cm, hence the difference between two DFT codes (different for only $\sim$ 30\cm) is larger than that between the DFT and eDMFT method. This is likely due to the basis set difference in the two DFT methods.

This is also consistent with our theoretical observation that a different magnetic order did not affect the relaxed crystal structure in the scheme of DFT+SOC+$U$ once the Ir $J_{\rm eff}$ = 1/2 state sets in.~\cite{Kim2016} However, it is also possible that the frustrated magnetism could play an important role in determining the highest $A_{\rm g}$ phonon energy (at about 587\cm~from the {\sc vasp} method) as this vibration is closely related to the local structure of Ir-O-Ir bond (as illustrated in Fig. 5d) in the main text), a key factor to determine exchange couplings.~\cite{Kim2016, Winter2016} Therefore, the origin of the mismatch of the highest $A_{\rm g}$ mode between the theory and experiment currently remains a topic for further investigation.

\begin{table}
\caption{Experimental and optimized structural information of \bLIO~at ambient pressure. The space group is $Fddd$ (No.~70, origin choice 2), where the internal coordinates for each inequivalent site are $(1/8, 1/8, z)$ for Ir and Li1/2, $(x, 1/8, 1/8)$ for O1, and $(x, y, z)$ for O2. In the DFT+SOC+$U$ calculation, cell parameters (${\bm a}$, ${\bm b}$ and ${\bm c}$) were allowed to change with the fixed volume, whereas in eDMFT fixed experimental cell parameters~\cite{Takayama:beta:Li213} were used. Ir-Ir and Ir-O bond lengths and Ir-O-Ir bond angles in each nearest neighbor bond are also given, where the Z- (X-) bonds denote Ir-Ir bonds parallel (not parallel) to along the $\bm{c}$-axis in Fig. 4a) in the main text. A zigzag-type antiferromagnetic and paramagnetic order was used for the DFT and eDMFT ($T$ = 232~K) calculation, respectively. Both DFT calculations in the table used both SOC and $U$.
}
\centering
\begin{tabular}{llrrrr}  \hline\hline
&             & Exp. & DFT  & DFT & eDMFT \\ [-3pt]
&             & (Ref.~\onlinecite{Takayama:beta:Li213}) & ({\sc vasp}) & ({\sc wien2k}) &  \\ \hline
                & $a$          & 5.910& 5.908    & 5.910 & 5.910 \\[-3pt]
                & $b$           &8.456& 8.440    & 8.456 & 8.456 \\[-3pt]
                 & $c$ (\AA) & 17.827 & 17.891  &17.827 &17.827 \\[3pt]
Ir   ($16g$)  & $z$ & 0.7085 & 0.7085 & 0.7096 & 0.7091  \\[3pt]
Li1 ($16g$)  & $z$ & 0.0498 & 0.0448 & 0.0460 & 0.0459 \\[3pt]
Li2 ($16g$)  & $z$ & 0.8695 & 0.8775 & 0.8783 & 0.8775 \\ [3pt]
O1 ($16e$)  & $x$ & 0.8572 & 0.8588 & 0.8614 & 0.8638 \\[3pt]
O2 ($32h$)  & $x$ & 0.6311 & 0.6320 & 0.6294 & 0.6277 \\[-3pt]
                      & $y$ & 0.3642 & 0.3654 & 0.3669 & 0.3666 \\[-3pt]
                      & $z$ & 0.0383 & 0.0384 & 0.0389 & 0.0393  \\ \hline
$d_{\rm Ir-Ir}$ & Z & 2.979 & 2.988 & 3.0203 & 3.000  \\[-3pt]
(in \AA)           & X & 2.973 & 2.973 & 2.9536 & 2.960  \\[3pt]
$d_{\rm Ir-O}$ & Z & 2.025 & 2.035 & 2.0573 & 2.059 \\[-3pt]
(averaged)      & X & 2.025 & 2.029 & 2.0356 & 2.043  \\[3pt]
$\theta_{\rm Ir-O-Ir}$ & Z & 94.68 & 94.50 & 94.45 & 93.50 \\[-3pt]
(degree)                     & X & 94.43 & 94.23 & 93.02 & 92.86 \\ \hline\hline
\end{tabular}
\label{tableA:str:ortho}
\end{table}

\begin{table}
\caption{Experimental and optimized lattice parameters and internal coordinates of \bLIO~with $C2/c$ (No.~15) space group symmetry at high pressure. In this calculation, lattice parameters of $\bm{a}$, $\bm{b}$, $\bm{c}$ and $\beta$ were optimized in the DFT and DFT+SOC+$U$ calculations with a fixed volume. Values of pressure measured in the experiment and DFT-estimation are shown in the top row. Ir-Ir and Ir-O bond lengths and Ir-O-Ir bond angles in each nearest neighbor bond are shown below.}
\centering
\begin{tabular}{llrrrr}  \hline\hline
&             & Exp. & DFT+SOC+$U$ &  DFT \\ [-3pt]
&             & (Ref.~\onlinecite{Veiga2017}) &   &   \\ \hline
& P (GPa) & 4.4 & 5.0 & 5.4 \\[3pt]
                & $a$           & 5.7930 & 5.7752 & 5.7485  \\[-3pt]
                & $b$           & 8.0824 & 8.0408 & 8.0319 \\[-3pt]
                 & $c$ (\AA) &  9.144 & 9.1951 & 9.2365 \\[3pt]
                 & $\beta$ (degree) & 106.777 & 106.263 & 106.016  \\[3pt]
Ir     & $x$ & 0.4219 & 0.4235 & 0.4238   \\[-3pt]
       & $y$ & 0.3844 & 0.3877 & 0.3887   \\[-3pt]
       & $z$ & 0.0780 & 0.0772 & 0.0770   \\[3pt]
Li1  & $x$ & 0.244 & 0.2434 & 0.2448  \\[-3pt]
       & $y$ & 0.632 & 0.6382 & 0.6401 \\[-3pt]
       & $z$ & 0.246 & 0.2436 & 0.2442 \\[3pt]
Li2  & $x$ & 0.926 & 0.9270 & 0.9261  \\[-3pt]
       & $y$ & 0.625 & 0.6177 & 0.6165 \\[-3pt]
       & $z$ & 0.589 & 0.5936 & 0.5932 \\[3pt]
O1  & $x$ & 0.7341 & 0.7320 & 0.7310 \\[-3pt]
       & $y$ & 0.3859 & 0.3895 & 0.3916 \\[-3pt]
       & $z$ & 0.2535 & 0.2542 & 0.2544 \\[3pt]
O2  & $x$ & 0.9024 & 0.9024 & 0.9015  \\[-3pt]
       & $y$ & 0.3598 & 0.3596 & 0.3585  \\[-3pt]
       & $z$ & 0.5792 & 0.5811 & 0.5812 \\[3pt]
O3  & $x$ & 0.4140 & 0.4118 & 0.4142 \\[-3pt]
       & $y$ & 0.3719 & 0.3625 & 0.3607 \\[-3pt]
       & $z$ & 0.5859 & 0.5870 & 0.5886 \\ \hline
$d_{\rm Ir-Ir}$ & dimer                   & 2.6609 & 2.5999 & 2.5838 \\[-3pt]
(in \AA)           & non-dimer            & 3.0136 & 3.0513 & 3.0697 \\[3pt]
$d_{\rm Ir-O}$ & dimer                  & 2.012 & 2.0098 & 2.0137 \\[-3pt]
(averaged)      & non-dimer           & 1.970 & 2.0235 & 2.0232 \\[3pt]
$\theta_{\rm Ir-O-Ir}$ & dimer        & 84.3 & 80.6 & 79.8 \\[-3pt]
(avg. deg.)             & non-dimer    & 97.4 & 97.9 & 98.7 \\ \hline\hline
\end{tabular}
\label{tableA:str:mono}
\end{table}

\section*{S6. Computational details}
\label{sm:calc}
\vspace{-0.2cm}

\subsection*{A. DFT+SOC+$U$ calculations}
\label{sm:calc:DFTU}
We employed {\sc vasp} to perform the electronic structure calculations, by using the projector-augmented wave basis set.~\cite{Kresse1996,Kresse1999} The same parameters for plane wave energy cutoff and k-point sampling used for the previous work~\cite{Kim2016} were chosen for the total energy and structural optimizations with experimental crystal structures at ambient~\cite{Takayama:beta:Li213} and high pressure.~\cite{Veiga2017} The calculations with and without including atomic SOC, the DFT+$U$ on-site Coulomb interaction,~\cite{Dudarev1998} and magnetism in the Ir $d$ orbital were done. All of calculations shown in this paper were done with the value of $U$ = 2 eV. We also checked that phonon spectra with $U$ = 2.5~eV showed similar results compared to the $U$ = 2~eV result (differences in frequencies smaller than 10\cm).

In the phonon calculation, we noticed that the lightest Li ions did not contribute high-energy modes significantly although it is the lightest ions, probably due to the much weaker ionic bonding with other ions. This makes sense because Li ions in Lithium-ion battery cathode materials are considered to be more freely removed than other constituent ions, as experimentally observed in \bLIO as well.~\cite{Pearce2017}

\subsection*{B. eDMFT calculations}
\label{sm:calc:eDMFT}
A fully charge-self-consistent DMFT method,~\cite{Haule2010} implemented in DFT + Embedded DMFT (eDMFT) Functional code,~\cite{eDMFT} which is combined with {\sc wien2k} code,~\cite{wien2k} was employed for computations of electronic properties and optimizations of internal coordinates.~\cite{Pascut2016} In DFT level the Perdew-Burke-Ernzerhof (PBE) generalized gradient approximation (GGA) is employed,~\cite{GGA-PBE} and different choices of the DFT exchange-correlation functional may affect quantitative natures of the results presented here. 2000 $k$-points were used to sample the first Brillouin zone with $RK_{\rm max}$ = 8.0. A force criterion of 10$^{-4}$ Ry/Bohr was adopted for optimizations of internal coordinates. A continuous-time quantum Monte Carlo method in the hybridization-expansion limit (CT-HYB) was used to solve the auxiliary quantum impurity problem,~\cite{HauleQMC} where the Ir $t_{\rm 2g}$ orbital was chosen as our correlated subspace in a single-site DMFT approximation. For the CT-HYB calculations, up to 10$^{10}$ Monte Carlo steps (at $T$ = 58 K) were employed for each Monte Carlo run.

In most runs, the temperature was set to be 232~K, but in some calculations with trial antiferromagnetic orders $T$ was lowered down to 58 K.  We tried to stabilize three different types of collinear antiferromagnetic orders (N\'{e}el-, zigzag-, and stripy-types), but all tried magnetic orders did not remain stable and the paramagnetic order still sets in down to $T$ = 58 K, which is rather unusual for the normal DMFT calculation as the DMFT result usually overestimates the ordering temperature, which may indicate an effect of the magnetic frustration of \bLIO.

The reasonable hybridization window of -10 to +10 eV (with respect to the Fermi level) was chosen, and $U$ = 5 eV and $J_{\rm H}$ = 0.8 eV of on-site Coulomb interaction parameters were used for the Ir $t_{\rm 2g}$ orbital. This values are slightly different compared to those used in another eDMFT study of layered perovskite and pyrochlore iridates,~\cite{ZhangPRL2013,ZhangPRL2017} $(U,J)$ = $(4.5, 0.8)$ eV, but this difference is not expected to lead to a qualitative difference.

Values of $U$ and $J_{\rm H}$ in eDMFT are significantly larger than those adopted in DFT+SOC+$U$ calculations because of the different choice of projectors for the correlated subspaces in both methods. For the Coulomb interactions, a simplified Ising-type (density-density terms only) approximation was applied to reduce the Monte Carlo noise, and a nominal double counting scheme was used with $n_d$ = 5 for the double counting correction.

Table~\ref{tableA:str:ortho} and \ref{tableA:str:mono} show the optimized crystal structures at ambient pressure (the orthorhombic structure) and high pressure (the monoclinic structure), starting from experimental structures, showing a reasonable agreement in both DFT+SOC+$U$ and eDMFT results.

\end{document}